# A Review of 3D Particle Tracking and Flow Diagnostics Using Digital Holography

Shyam Kumar M[1] and Jiarong Hong[1,2,*]

[1]University of Minnesota, Department of Mechanical Engineering, Minneapolis, MN, USA

[2]Saint Anthony Falls Laboratory, University of Minnesota, Minneapolis, MN, USA

*jhong@umn.edu

**Abstract:** Advanced three-dimensional (3D) tracking methods are essential for studying particle dynamics across a wide range of complex systems, including multiphase flows, environmental and atmospheric sciences, colloidal science, biological and medical research, and industrial manufacturing processes. This review provides a comprehensive summary of 3D particle tracking and flow diagnostics using Digital Holography (DH). We begin by introducing the principles of DH, accompanied by a detailed discussion on numerical reconstruction. The review then explores various hardware setups used in DH, including inline, off-axis, and dual or multiple-view configurations, outlining their advantages and limitations. We also delve into different hologram processing methods, categorized into traditional multi-step, inverse, and machine learning-based approaches, providing in-depth insights into their applications for 3D particle tracking and flow diagnostics across multiple studies. The review concludes with a discussion on future prospects, emphasizing the significant role of machine learning in enabling accurate DH-based particle tracking and flow diagnostic techniques across diverse fields, such as manufacturing, environmental monitoring, and biological sciences.

## 1. Introduction

Particle tracking is a fundamental tool in both flow diagnostics and the study of particle dynamics, offering crucial insights across a broad spectrum of scientific and engineering applications. In flow diagnostics, particle tracking velocimetry (PTV) has emerged as an indispensable technique for investigating fluid behavior across scales ranging from microfluidic devices to turbulent flows [1]. For instance, in microfluidics, the ability to precisely track particles in three dimensions is critical for optimizing device performance, enabling advances in lab-on-a-chip technologies and other microscale systems [2]. Similarly, in the study of turbulence, the necessity of three-dimensional (3D) tracking becomes even more pronounced. Capturing the full 3D velocity field is essential for reconstructing the velocity gradient tensor, which plays a key role in understanding pressure distributions and the intricate dynamics inherent in turbulent flows [3].

Beyond flow diagnostics, advanced 3D tracking methods are essential in studying particle dynamics across a diverse range of complex systems, including multiphase flows, environmental and atmospheric sciences, colloidal science, biological and medical research, and industrial manufacturing processes. In multiphase flows, tracking the movement of droplets and bubbles in turbulent fields is crucial for optimizing processes like spray atomization in fuel injectors and chemical reactors [4,5]. In environmental and atmospheric sciences, accurate 3D tracking is essential for understanding cloud microphysics and pollutant dispersion, which are fundamental to developing precise climate models and assessing environmental health [6-11]. Similarly, in colloidal science, understanding particle dynamics is critical for controlling the behavior, stability, and functionality of colloidal systems in various scientific and industrial applications [12,13]. In biological and medical research, the 3D motion of microorganisms plays a central role in phenomena such as biofilm formation, predator-prey interactions, and sperm motility, which have significant implications for health and disease [14-16]. In industrial manufacturing processes, precise 3D tracking is critical for optimizing operations such as powder metallurgy and spray drying, leading to improved product quality and efficiency [17].

The increasing complexity and 3D nature of these processes highlight the limitations of traditional two-dimensional (2D) tracking methods, which, while useful, often fail to capture the full scope of interactions that occur in real-world systems. As scientific inquiry and technological innovation continue to advance, the ability to accurately track and analyze particles in 3D becomes not just advantageous but essential. The transition to 3D particle tracking represents a critical leap forward, enabling researchers and engineers to fully capture the spatial complexity of flow fields and particle behaviors. Commonly used 3D particle tracking techniques, such as tomographic PTV, defocusing, synthetic aperture, and plenoptic imaging, typically rely on multiple cameras or segmented sensors to reconstruct 3D particle fields in larger sample volumes [18]. These methods have been widely adopted across various applications but often require complex setups and are best suited for larger-scale measurements.

In contrast, digital holography (DH) offers a versatile and efficient approach to 3D particle tracking by capturing detailed volumetric information through interference patterns using a single imaging sensor. This capability provides high spatial and temporal resolution along with a significant depth of field, making DH a powerful tool across a wide range of disciplines. While the depth of field in conventional photography and microscopy is defined as the range over which particles appear acceptably sharp, in holography this concept differs fundamentally. In holography, the diffraction patterns of particles are recorded as 2D holograms on the camera sensor. During the reconstruction step, particles can be numerically retrieved provided their diffraction patterns reach the sensor and are adequately captured in the recorded hologram [19]. The effective depth of field in holography depends on the farthest particle from the camera whose diffraction pattern is recorded. While magnification, aperture size, and other optical parameters affect the depth of field in both conventional photography and microscopy, as well as in holography, additional holography-specific factors—such as laser power and coherence, the scattering characteristics of the particles (e.g., morphology, refractive index, and medium), optical system noise, cross-interference of diffraction patterns, and the chosen numerical reconstruction algorithm—further influence the achievable depth of field [20]. Notably, the depth of field in holography is approximately three orders of magnitude larger than that in conventional photography and microscopy [21]. For instance, at a magnification of 10X, the depth of field in brightfield microscopy is about 10 μm, whereas in digital holographic microscopy it extends to about 10 mm [21]. Similarly, while imaging spray droplets using shadowgraphy with about 1X magnification is often limited to a few millimeters, in holography the depth of field can extend to nearly a meter, depending on the system configuration and the scattering properties of the particles.

In fluid mechanics and flow diagnostics, DH has been utilized to achieve precise 3D measurements in various flow regimes. In microfluidic systems, where small-scale flows demand accurate tracking of particle dynamics, DH provides high-resolution data essential for device optimization and understanding microscale phenomena [22]. Separately, in the study of wall-bounded turbulence, DH has enabled high-resolution imaging of 3D structures within turbulent boundary layers, such as the buffer layer [23] and the viscous sublayer [24]. This has provided deeper insights into turbulent flow behavior that are unattainable with 2D methods, enhancing our understanding of turbulence mechanics. In biomedical and life sciences, DH has been instrumental in tracking the complex 3D swimming patterns of microorganisms [25,26] and studying sperm motility, which has critical implications for reproductive health [27,28]. In industrial processes and chemical engineering, DH has been applied to monitor the dynamics within fluidized beds, offering valuable data for optimizing processing conditions [29]. Furthermore, in environmental and atmospheric sciences, DH has shown potential in tracking particles for pollutant dispersion modeling [30] and monitoring aerosol behaviors, which are vital for air quality assessments and climate studies. In agricultural sciences, DH has been used for monitoring the 3D movement of pests or beneficial organisms in crop fields [31,32], contributing to more efficient and sustainable farming practices. These diverse applications across various fields underscore the broad appeal and versatility of DH as a tool for 3D particle tracking, enabling advancements in both fundamental research and practical applications.

Although there are extensive reviews on 3D particle tracking [18], the application of holography in fluid mechanics [19] and multi-phase flows [33], a focused review on particle tracking and flow diagnostics across different fields specifically using digital holography is still lacking. This gap is increasingly important to address, given the evolution of diverse methods and advanced data processing techniques that have been developed to accurately visualize flow fields and track particles using DH. While Memmolo et al. (2015) provides valuable insights into the development of holographic 3D particle tracking, particularly focusing on bio-microfluidics, it places less emphasis on other application areas and does not extensively cover more recent data processing techniques, such as those involving machine learning (ML) [34]. Unlike tomographic PTV, which typically employs a Eulerian framework based on cross-correlation for flow analysis, DH tracking operates within a Lagrangian framework, requiring the precise detection and localization of individual particles. This review aims to fill that gap by concentrating on the critical techniques for particle detection and localization within DH, including the latest advancements in ML approaches. Additionally, we will examine how these techniques are applied across various fields, demonstrating the broad impact and potential of DH in both research and industry. In Section 2, the principles of digital holography are presented. Details of different holography setups are provided in Section 3. Various methods developed for hologram processing are reviewed in Section 4, followed by a detailed summary, current limitations and challenges, and future prospects in Sections 5 to 7.

## 2. Principle of digital holography

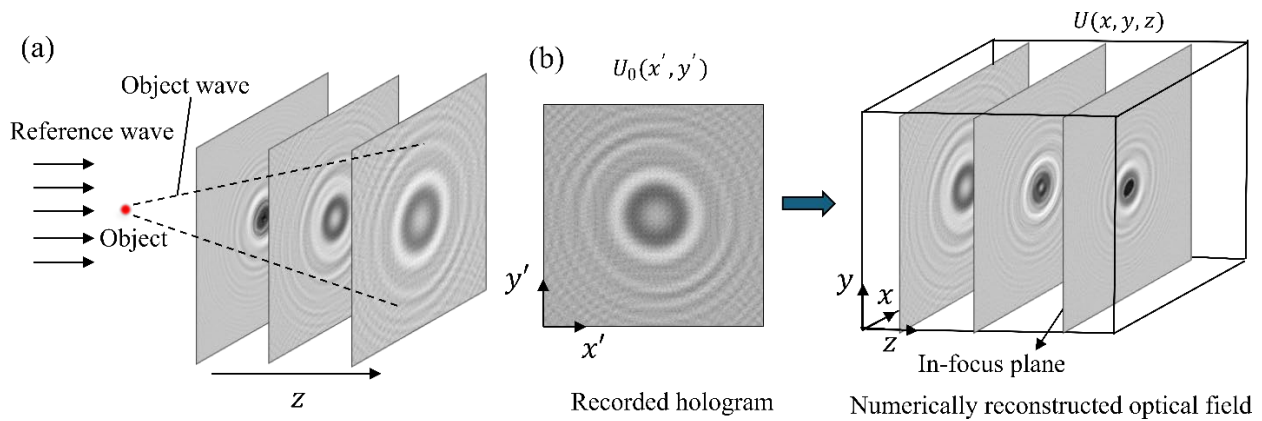

**Fig. 1.** Schematics showing (a) the hologram formation by the interference between the reference wave and the object wave, and (b) the corresponding numerical reconstruction of the recorded hologram.

Digital holography is a computational imaging technique based on the principle of interference between light waves, enabling the capture of both amplitude and phase information of an object [35]. In DH, a coherent light source, typically a laser, illuminates an object, resulting in scattered light known as the object wave $O(r)$ (Fig. 1a). This object wave carries detailed information about the object's physical characteristics, such as size, shape, and refractive index variations. At the same time, a reference wave $R(r)$, usually a coherent plane wave, is directed toward the sensor (Fig. 1a). The interference between the object and reference waves produces a hologram, which is captured as an intensity distribution $I_{holo}$ on a digital sensor, such as a CCD or CMOS camera. The recorded intensity can be expressed as:

$$I_{holo}(r) \approx |R|^2 + |O|^2 + R \cdot O^*(r) + R^* \cdot O(r) \quad (1)$$

where, $I_{holo}(r)$ represents the hologram, or the recorded intensity pattern, $|R|^2$ and $|O|^2$ is the intensity of the reference wave and object wave, respectively, $R \cdot O^*(r)$ represents the interference between the reference light and the complex conjugate of the object wave, and $R^* \cdot O(r)$ accounts for the conjugate terms. Since the object wave $O(r)$ often exhibits spherical wavefronts while the reference wave $R(r)$ typically has planar wavefronts, their interference results in characteristic concentric fringes in the hologram [36]. The recorded intensity thus contains both amplitude and phase information of the object, although the phase information is indirectly captured within the interference pattern.

After recording the hologram, computational techniques are employed to reconstruct the original object wave and recover the 3D distribution of the particles [37]. This is achieved by numerically propagating the recorded wave back into space, a process commonly referred to as digital reconstruction (Fig. 1b). This involves calculating both the amplitude and phase of the light waves that interacted with the object, allowing for a detailed representation of the object's physical characteristics. Phase retrieval is crucial in many applications, such as measuring optical path length variations and reconstructing objects located at different depths. Digital holography offers an extended depth of field, enabling the simultaneous focusing of objects located at different axial positions from a single recorded hologram. This is a significant advantage in applications where multiple particles are distributed across a large volume, as it allows for the reconstruction of the entire particle field without the need for multiple images. However, the resolution of the reconstructed images depends on various factors, including the wavelength of the illumination, the distance between the object and the sensor, and the numerical aperture of the optical system.

The basic equation for holographic reconstruction is derived from the principles of wave propagation and diffraction. A commonly used starting point is the scalar diffraction integral, which describes how a wavefront propagates from one plane to another:

$$U(x, y, z) = U_0(x', y') * h(x', y', z) \quad (2)$$

where $U$ is the reconstructed complex optical field a distance $z$ from the hologram plane (Fig. 1b), $U_0$ the recorded hologram intensity on the sensor plane, $h$ is the diffraction kernel, describing how light diffracts as it propagates from the hologram plane to a reconstruction plane at a distance z and * denotes the convolution operation, $x$, $y$ and $z$ are lateral and longitudinal locations respectively, $x'$, $y'$ represents the lateral locations in the imaging plane (camera sensor). Several numerical methods have been developed for reconstructing holograms, each based on different diffraction models or approximations. Below are some commonly used reconstruction techniques along with their corresponding equations:

**The Rayleigh-Sommerfeld diffraction** theory provides a rigorous solution for wave propagation without making paraxial or far-field approximations and is suitable for reconstructing fields from objects with irregular or unknown shapes [35]. The reconstruction formula is given by:

$$U(x,y,z) = \frac{1}{i\lambda} \iint U_0(x',y') \frac{e^{ikr}}{r} \left(\frac{z}{r}\right) dx'dy' \qquad (3)$$

where $r = \sqrt{(x-x')^2 + (y-y')^2 + z^2}$ is the distance between a point $(x',y')$ on the hologram and a point $(x,y)$ at the reconstruction plane, λ is the wavelength of the light and $k = \frac{2\Pi}{\lambda}$ is the wave number. This model fully accounts for the spherical nature of wave propagation from each point on the hologram.

**The Fresnel approximation** simplifies the Rayleigh-Sommerfeld theory by assuming that the propagation distance z is much larger than the wavelength λ and that the angles involved are small (paraxial approximation) [35]. The corresponding reconstruction formula is:

$$U(x,y,z) = \frac{e^{ikz}}{i\lambda z} \iint U_0(x',y') e^{\frac{ik}{2z}((x-x')^2 + (y-y')^2)} dx'dy' \qquad (4)$$

Using the simplified diffraction kernel improves the computational speed and is suitable when the object is far from the hologram plane. However, the accuracy degrades for objects closer to the hologram.

**The angular spectrum method** reconstructs the hologram by decomposing the wavefield into a spectrum of plane waves using Fourier transforms [38,39]. The propagation is calculated in the frequency domain:

$$U(x,y,z) = \mathcal{F}^{-1}\{\mathcal{F}[U_0(x,y)] \cdot H(f_x, f_y, z)\} \qquad (5)$$

where $\mathcal{F}$ and $\mathcal{F}^{-1}$ are the Fourier and inverse Fourier transforms, $H(f_x, f_y, z) = e^{ikz\sqrt{1-(\lambda f_x)^2 - (\lambda f_y)^2}}$ is the transfer function with $f_x$ and $f_y$ being the spatial frequency coordinates. This diffraction theory is accurate for both near-field and far-field conditions and facilitates easy manipulation such as filtering and aberration correction. However, the assumption of planar wave components may introduce less accurate reconstructions for highly curved wavefronts, and this method requires adequate sampling to avoid artifacts due to aliasing.

**Deconvolution methods** have been introduced to improve the accuracy of holographic reconstructions by addressing the blurring inherent in diffraction processes [40-42]. These methods treat the blurring as a convolution of the true object field with the diffraction kernel and aim to reverse this process through mathematical algorithms. The deconvolved (reconstructed) complex field at distance z is:

$$U(x,y,z) = \mathcal{F}^{-1}\left\{\frac{\mathcal{F}[U_0(x,y)]}{\mathcal{F}[h(x,y,z)] + \epsilon}\right\} \qquad (6)$$

where $\epsilon$ is a regularization parameter to mitigate division by zero and reduce noise amplification. Deconvolution improves the resolution of reconstructions, making it possible to distinguish fine details in the 3D optical field. While deconvolution techniques can partially mitigate noise and interference, particularly when advanced algorithms like iterative optimization are employed, they inherently introduce artifacts and depend strongly on accurately modeling the diffraction kernel. Poorly defined kernels or dynamic noise in the optical system may lead to reduced localization of out-of-focus signals, residual artifacts, negative intensities, depth elongation causing axial blur, edge artifacts, or aliasing in reconstructed images [40,42,43]. In addition, different deconvolution approaches, such as Wiener filtering and iterative methods, can introduce their own characteristic distortions, including ringing near high-contrast edges, over-smoothing, noise amplification in low-signal regions, convergence-related artifacts, and spurious patterns arising from overfitting during iterative optimization [44]. Consequently, achieving precise diffraction kernel modeling and implementing balanced regularization strategies are imperative for minimizing noise while maintaining high image fidelity.

Fast Fourier Transform (FFT) algorithms are extensively employed to expedite the reconstruction process in DH. By operating in the frequency domain, the convolution in the reconstruction equation (Eq. 2) is reduced to a simple point-wise multiplication, substantially decreasing computational overhead. This acceleration is critically important in real-time or high-resolution hologram analysis, where computational efficiency directly influences the feasibility and quality of the final results.

## 3. Hardware setup of digital holography

Digital holography encompasses various hardware configurations, including inline, off-axis, and specialized setups, each designed to address specific imaging challenges and application requirements. The choice of configuration plays a pivotal role in optimizing the balance among simplicity, accuracy, spatial resolution, and the nature of the information extracted from the hologram.

### 3.1. Digital Inline Holography

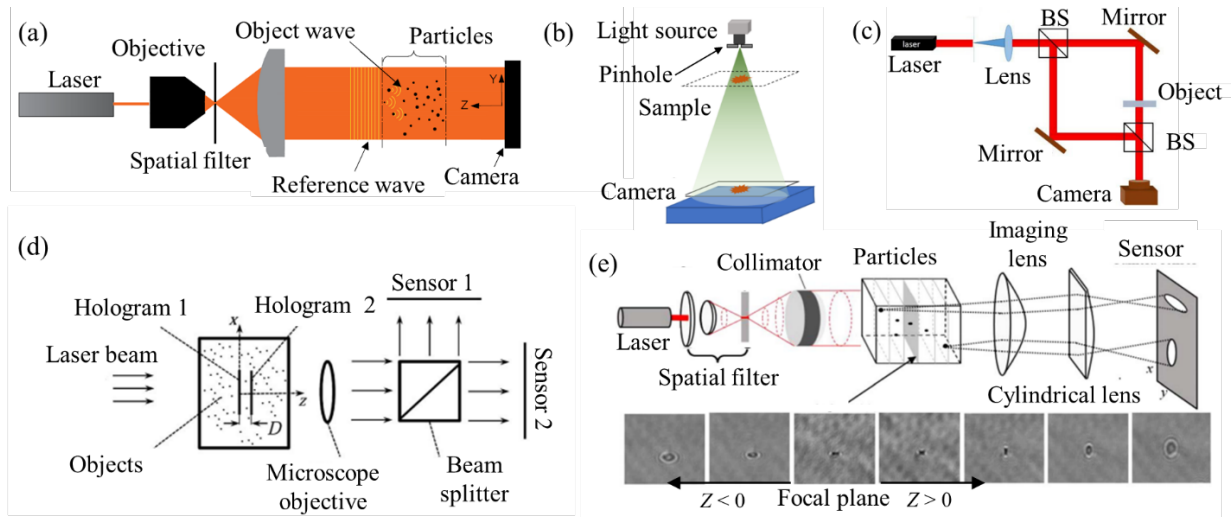

**Fig. 2.** Schematic of optical arrangement of different types of DIH including (a) collimated setup, adapted from [49] Copyright (2024), with permission from Elsevier, (b) lensless microscopy setup, adapted from [51], (c) Mach–Zehnder setup, adapted from [51], and setups to resolve twin image problem via (d) two inline DIH with focal planes separated by a short distance, adapted with permission from [62] © Optical Society of America and (e) astigmatic effect using cylindrical lens, adapted from [63].

Digital Inline Holography (DIH) is one of the simplest and most widely utilized configurations in DH [45]. Its appeal lies in its straightforward optical setup, which requires minimal and cost-effective components and offers simplified image calibration compared to other holographic techniques. DIH is particularly advantageous for miniaturized optical sensing applications and environments where physical access to the sample volume is constrained. In a typical DIH setup, a collimated light beam is directed through the sample volume, where particles or objects scatter the light. The scattered light, known as the object wave, interferes with the unscattered portion of the beam serving as the reference wave. This interference produces a hologram that is recorded on a digital sensor, such as a CCD or CMOS camera (Fig. 2a). The simplicity of this arrangement enables the capture of particles distributed throughout the entire sample volume, making DIH a versatile tool for a wide range of applications.

DIH is highly valued for its simplicity and effectiveness in capturing particle distributions without the need for complex optical components. This straightforward setup has led to its widespread use in flow diagnostics and particle detection across various applications. For example, DIH has been successfully employed to study aerosol dynamics [30,46], investigate water droplets [47-49], analyze bubbles, and explore complex environments like fluidized beds [29]. DIH has been utilized to observe microorganisms [25] and detect particles in coal research [50]. Furthermore, DIH has been adapted for studying 3D structures and motion in biological samples. By implementing a lensless configuration with a point-source reference wave, generated by passing light from an LED through a pinhole (Fig. 2b), DIH achieves high-resolution imaging where magnification is determined by the distance between the reference source and the object [51]. This setup allows researchers to resolve fine structural details, making DIH valuable for examining individual cells and microorganisms [52-55].

Despite these advantages, inline holography faces certain limitations that affect its precision in some applications. A significant drawback is its reduced longitudinal resolution, which limits the accuracy of measuring particle positions along the optical axis [56]. Another challenge is the presence of twin images, where the real and virtual images of the particle field overlap and distort the reconstructed image [57,58]. Additionally, noise interference can degrade the quality of recorded holograms, complicating the accurate reconstruction of the particle field. To address these issues, several techniques have been developed, although they may not fully resolve all limitations of DIH. The Mach-Zehnder method (Fig. 2c) introduces a reference wave undisturbed by the object field to reduce noise in the hologram, thereby enhancing the signal-to-noise ratio [51,59-61]. In this configuration, the reference and object beams are separated and then recombined, improving the clarity of the holographic image. Ling and Katz (2014) tackled the twin image problem by recording two parallel inline holograms with a slight separation between their initial focal planes (Fig. 2d) [62]. During reconstruction, the real images from both holograms overlap while the twin images are displaced. By correlating the spatial intensity distributions, real and twin images can be effectively distinguished, enhancing image accuracy. Zhou et al. (2020) employed a cylindrical lens (Fig. 2e) to induce astigmatism in the holographic system [63]. This approach alters the elongation pattern of holograms based on the particle's position relative to the focal plane: particles closer to the laser source exhibit horizontal elongation, while those nearer to the camera show vertical elongation. This distinction helps resolve twin image artifacts and accurately determine the focal plane. However, even with these methods, the inherent limitations of inline holography in precisely measuring particle positions along the optical axis persist, particularly in applications requiring high depth resolution. To overcome these challenges, off-axis holography has emerged as a more suitable technique. Off-axis holography provides better separation of real and twin images and enhances depth resolution, making it preferable for applications that demand higher precision in longitudinal measurements.

## 3.2. Off-axis Holography

In off-axis holography, the reference wave is tilted relative to the object wave by a certain angle, resulting in spatial modulation of the interference pattern recorded in the hologram (Fig. 3a and 3b). This angular separation is crucial for shifting the interference terms in the Fourier domain, allowing the real image, twin image, and zero-order (DC) terms to be spatially separated. Such separation is essential for extracting the real image without contamination from unwanted artifacts like the twin image or the zero-order term [36]. Eq. (1) can be modified and the intensity of the hologram recorded in off-axis holography can be described mathematically as:

$$I_{holo}(r) \approx |R|^2 + |O|^2 + R\,e^{i2\pi f_r \cdot r} \cdot O^*(r) + R^* e^{-i2\pi f_r \cdot r} \cdot O(r) \quad (7)$$

where $f_r$ is the spatial frequency shift introduced by the angle between the reference and the object waves, $e^{i2\pi f_r \cdot r}$ represents the phase modulation due to the angular separation, $R\,e^{i2\pi f_r \cdot r} \cdot O^*(r)$ is the interference term that corresponds to the real image of the object. This term is spatially shifted in the Fourier domain due to the phase modulation from the reference beam. The twin image, which is generated by the conjugate interference term $R^* e^{-i2\pi f_r \cdot r} \cdot O(r)$, is also shifted in the opposite direction in the Fourier domain. By eliminating the overlap between the real and twin images through spatial modulation, the angular separation enhances the longitudinal resolution in the reconstructed image compared to DIH. As a result, this technique is well-suited for applications requiring high longitudinal resolution in 3D reconstructions and precise quantitative analysis, making it valuable for scientific research and industrial uses where accuracy and artifact reduction are essential. Its ability to prevent blockage of the reference wave by particles in dense environments makes it particularly suitable for diagnostics involving high particle concentrations, especially in holography particle image velocimetry (PIV) [64] (Fig. 3a). Moreover, off-axis holography is also utilized in holographic microscopy (Fig. 3b), enabling high-accuracy 3D imaging of biological samples in microfluidic systems [65-68].

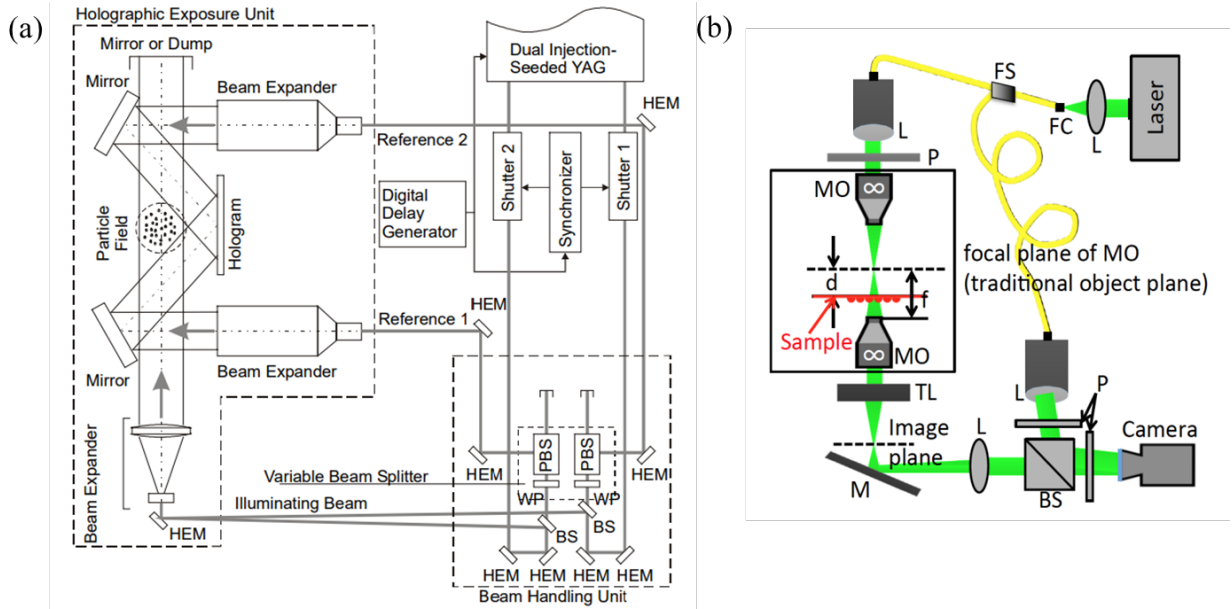

**Fig. 3.** Schematic of optical arrangement of off-axis setup used for (a) holography PIV adapted from [64] Copyright (2024), with permission from Springer Nature (b) holographic microscopy adapted from [68].

Although off-axis holography offers significant benefits, it also has certain limitations. A major drawback is the requirement for a stable and powerful laser source with a long coherence length to maintain precise phase relationships. This necessity, along with the additional optical components needed compared to DIH, increases both the complexity and cost of the system. Furthermore, its complex setup is highly sensitive to misalignment. Even slight errors can result in reduced fringe contrast, unintended phase shifts, and optical aberrations, which collectively degrade the quality of the reconstructed images.

### 3.3. Dual-view or multiple-view holography

Dual-view or multiple-view holography offer an alternative to off-axis holography for achieving high longitudinal resolution (Fig. 4). These techniques involve capturing holograms from two or more different perspectives relative to the object of interest. By integrating information from multiple angles, they provide a more comprehensive and accurate reconstruction of the object's 3D position. These methods are particularly advantageous in situations where capturing the intricate 3D nature of flow or particle dynamics is essential. Furthermore, in environments with high particle concentrations, significant positional errors can occur due to increased cross-interference noise. The overlapping diffraction patterns from densely packed particles complicate the accurate reconstruction of individual particle positions. Dual-view or multiple-view holography address this issue by capturing holograms from multiple perspectives, which helps to reduce cross-interference noise. By integrating information from different angles, these methods enhance the accuracy of particle position measurements.

To reduce complexity in the experimental setup, Sheng et al. (2003) developed a single-beam dual-view holographic PIV system that utilizes a mirror placed in the test section at a 45° angle (Fig. 4a) [69]. This arrangement allows particles in the illuminated volume to be captured from two different directions. Both the direct and reflected views are recorded on the same hologram. During reconstruction, these views are simultaneously reconstructed but at different locations in space. By combining the elongated particle traces from both views, researchers can accurately determine the exact 3D locations of the particles. This method was applied to measure 3D wake flow of rising bubbles. Similarly, Gao and Katz (2018) employed dual-view tomographic holographic PIV involving two cameras to measure 3D coherent flow structures around a cubic roughness element embedded in the inner part of a high Reynolds number turbulent boundary layer (Fig. 4b) [70]. In such scenarios, the flow field is dominated by complex interactions involving vortices shed from the roughness, and dual-view holography enables detailed characterization of these dynamics.

Despite the enhanced imaging capabilities offered by dual or multiple-view holography, these techniques come with certain limitations. The necessity of employing multiple holographic recording systems adds complexity to the experimental setup and increases costs due to the additional equipment required. Furthermore, merging and processing data from several viewpoints require greater computational power and sophisticated algorithms, leading to higher data processing overhead and storage demands. The techniques are also sensitive to distortions; misalignments or aberrations in any of the views can adversely affect the overall reconstruction quality. Nonetheless, in situations where utmost accuracy and resolution are critical, especially in turbulent flows and environments with high particle concentrations, dual or multiple-view holography offer substantial benefits.

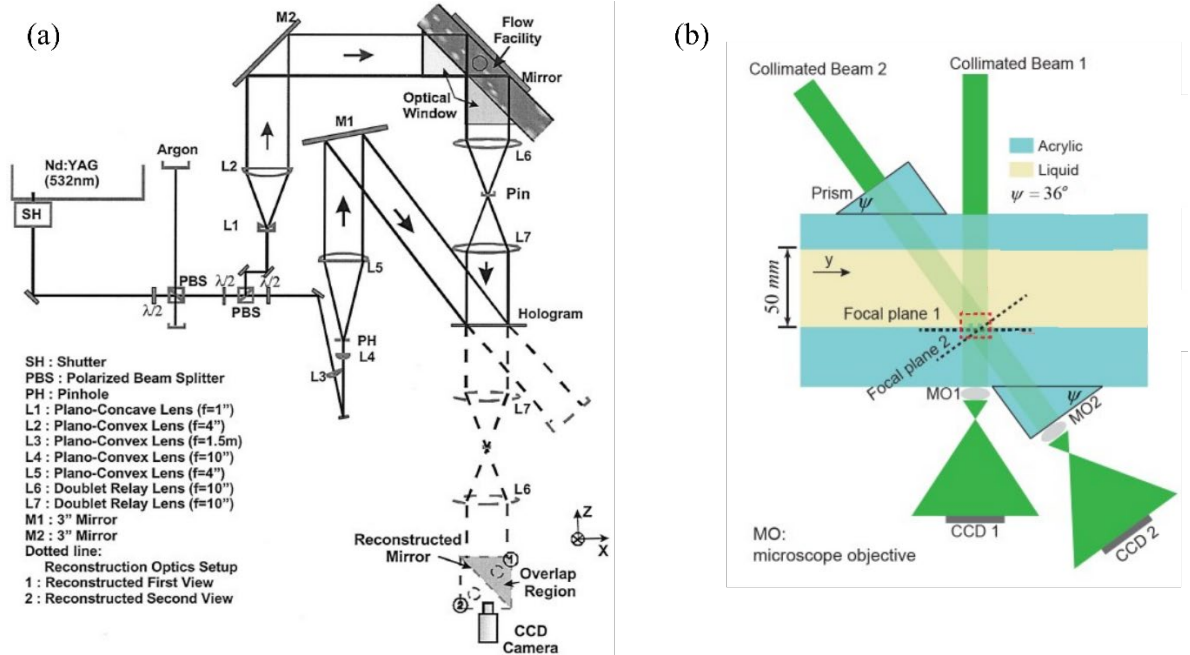

**Fig. 4.** (a) Single-beam dual-view holographic PIV system for measuring 3D flow in the wake of a rising bubble. Adapted with permission from [69] © Optical Society of America. (b) Dual-view tomographic holographic PIV for measuring 3D coherent flow structures around a cubic roughness element embedded in the inner part of a high Reynolds number turbulent boundary layer. Adapted with permission from [70] © Optical Society of America.

## 4. Hologram data processing for 3D particle tracking

Holograms encode 3D particle field information into 2D interference patterns. Extracting 3D particle trajectories from these recorded holograms is the most challenging aspect of applying DH to flow measurements and particle tracking. In the literature, various hologram processing methods have been developed, which can be broadly categorized into traditional multi-step methods, inverse methods, and ML approaches. Each category offers distinct advantages in terms of computational efficiency, accuracy, and applicability to complex flow scenarios, for example, cases with high cross interference (high particle concentration flows, polydisperse particle fields), wide ranges of particle depth, or particle fields with complex morphologies and optical properties. In this section, we provide a detailed overview of these methods.

### 4.1. Traditional multi-step approach

The traditional multi-step approach to holographic data processing entails two primary steps. First, the recorded hologram is numerically reconstructed at multiple planes, yielding a 3D optical field. Second, the location of each particle within this 3D field is determined. There are generally two ways to localize particles following reconstruction. In the first approach, longitudinal focus metrics identify the in-focus plane for each particle [19,33,34], and 2D segmentation (e.g., adaptive thresholding or edge-based detection) establishes the particle's lateral center [71-75]. In the second approach, a full 3D segmentation is applied directly to the reconstructed volume to detect and localize particles in three dimensions. This second method can be more robust in dense particle fields but often entails higher computational costs.

During the reconstruction step, most multi-step methods rely on diffraction-based models or approximations (see Section 2) to propagate the optical field and generate a 3D representation. Convolution with a diffraction kernel is commonly used but can suffer from a low signal-to-noise ratio (SNR) in high particle concentration environments. Deconvolution methods have been proposed to improve SNR, enhance longitudinal resolution, and suppress out-of-focus noise. Toloui and Hong (2015) pioneered the use of deconvolution-based reconstruction for 3D flow measurements in microfluidic channels, demonstrating improved clarity and accuracy in high particle density regions [42]. Their subsequent work (Toloui et al. 2017) extended this technique to smooth-wall turbulent flows [76], while Toloui et al. (2019) applied deconvolution-based methods to quantify flow-structure interactions in turbulent flows over flexible rough surfaces [77]. Despite these advancements, the higher computational load can limit the feasibility of deconvolution for real-time or large-scale implementations. Although deconvolution can yield more precise particle reconstructions, it is computationally intensive and sensitive to the accuracy of the underlying kernel.

In the subsequent particle localization step, the first approach determines each particle's longitudinal (depth) position by applying focus metrics, which can be loosely classified into three main categories: spatial-based, spectral-based, and wavelet-based methods. Specifically, the spatial-based focus metrics operate directly on reconstructed images to quantify focus levels. Edge detection-based methods, such as those proposed by Tian et al. (2010), use thresholded edge detectors to identify the focused plane within a sequence of reconstructed images at different depths [78]. Sharpness metrics, including the Laplacian sharpness metric, assess focus by calculating the image plane with maximum sharpness, often using measures like the variance of the Laplacian (Fig. 5a) [79-82]. For instance, Choi et al. (2012) applied sharpness-based focus metrics to achieve 3D tracking of free-swimming phytoplankton (Fig. 6a) [83]. The Tenengrad operator [84-86], another widely used focus metric, computes gradient magnitudes using the Sobel operator [87] in both horizontal and vertical directions, summing the squared gradients to quantify sharpness (Fig. 5b). Its advantage lies in its robustness to noise, as it utilizes multiple gradient samples per pixel. Conversely, the Brenner gradient method [88] calculates sharpness based on intensity differences between neighboring pixels. While computationally efficient, it is more sensitive to noise due to its reliance on only two sampling points per pixel. Other spatial domain methods include the entropy method [89], which measures image entropy to assess focus, and image correlation techniques [90], which compare reconstructed images at different depths. Additionally, constrained least squares filtering [91] has been employed to enhance focus detection. However, spatial domain focus metrics can degrade significantly in noisy images, especially in holograms with high particle concentration where cross-interference occurs. Their effectiveness depends heavily on the quality of image edges, which may be compromised by optical distortions.

Spectral-based focus metrics, introduced by Li et al. (2007), operate in the frequency domain [92]. Specifically, the L1-norm method uses the L1-norm to measure the sparsity of spectral components. A sparser spectrum indicates a better-focused image. This method has been applied by Kumar et al. (2016) to monitor Drosophila in 3D, extracting various behavioral parameters (Fig. 6b) [93]. By utilizing this approach, they were able to image complex leg and wing motions of flies at a resolution that captures specific landing responses. Instead of the L1-norm, Arias-Sosa et al. (2024) employed logarithmic weighting in the spectral domain (Fig.5c), providing a potentially more stable focus measure than the L1-norm under noisy or complex conditions [94]. Spectral methods can be computationally efficient since they process frequency domain data directly, avoiding the need for reconstructing images at multiple depths. However, spectral methods may struggle when noise or cross-interference distorts the spectral content, leading to inaccuracies in focus detection [95]. They may also be less effective with complex particle morphologies or at higher concentrations.

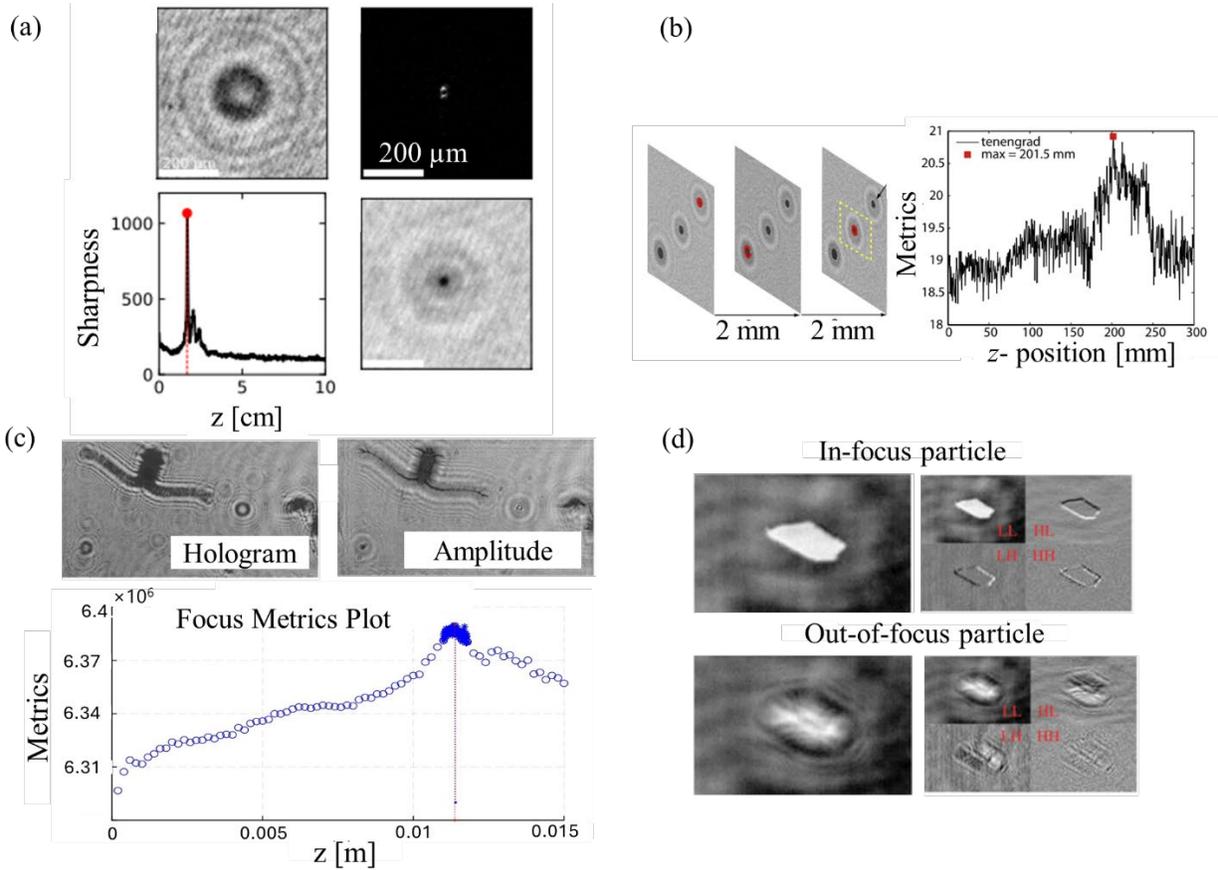

**Fig. 5.** Examples of holograms and the longitudinal variation of pixel intensity at different reconstruction planes, illustrating different focusing metrics used in digital holography. (a) The top panel shows the experimentally recorded hologram of a water droplet and the corresponding spatial map of droplet sharpness near the plane of best focus. The bottom panel depicts the variation of mean droplet sharpness with longitudinal locations (z) and the plane reconstructed at maximum sharpness, marked in red on the sharpness vs. z plot. Adapted from [82]. (b) Application of the Tenengrad operator on a synthetic hologram. Reconstructions at three different z-locations are shown, along with the Tenengrad metric variation with z for the particle highlighted in the yellow rectangle. Adapted from [85]. (c) Focus detection of planktonic organisms using logarithmic weighting in the spectral domain. The top panel displays the hologram of planktonic organisms and the reconstructed image at the in-focus plane determined using the focus metric. The bottom panel shows the variation of the focus metric with the z-location. Adapted from [94]. (d) Wavelet-based focusing method applied to a coal particle image. The reconstructed image and the corresponding decomposed high- and low-frequency subimages are shown for the in-focus (upper panel) and out-of-focus (middle panel) cases. The variance of the image gradient in the subimages is used as a focus metric to extract the focus plane. Adapted with permission from [96] © Optical Society of America.

Wavelet-based focus metrics, proposed by Wu et al. (2014), decompose the reconstructed image into high- and low-frequency subimages using wavelet transforms (Fig. 5d) [96]. By analyzing the variance of the image gradient in the high frequency subimages, the focus plane can be accurately determined. Wavelet-based methods allow for detailed local analysis and offer robustness to noise, making them effective even in complex holographic reconstructions, such as dealing with both opaque and transparent objects, noisy environments, and handling high-concentration particle fields. However, these methods can be computationally intensive due to the need for wavelet decomposition and reconstruction at multiple scales.

The second particle localization approach utilizes 3D segmentation for holographic particle image velocimetry (HPIV), making it particularly well-suited for dense particle fields. This technique is based on 3D blob analysis, where reconstructed particle traces are initially segmented into 2D planes using intensity-based thresholding. Adjacent 2D particle segments at different depths are then connected through a continuity operator to form coherent 3D blobs. The center of mass, or intensity-weighted centroid, of these blobs is subsequently computed to determine accurate 3D particle positions (Sheng et al., 2009) [97]. This method has been effectively applied to measure 3D turbulent flow with significantly higher resolution compared to tomographic PIV and PTV, in both smooth-wall (Sheng et al., 2009) and rough-wall environments (Talapatra and Katz, 2012) [97, 98]. Despite its strengths in handling high particle concentrations, this localization strategy is computationally intensive and faces challenges when cross-interference among particles becomes significant.

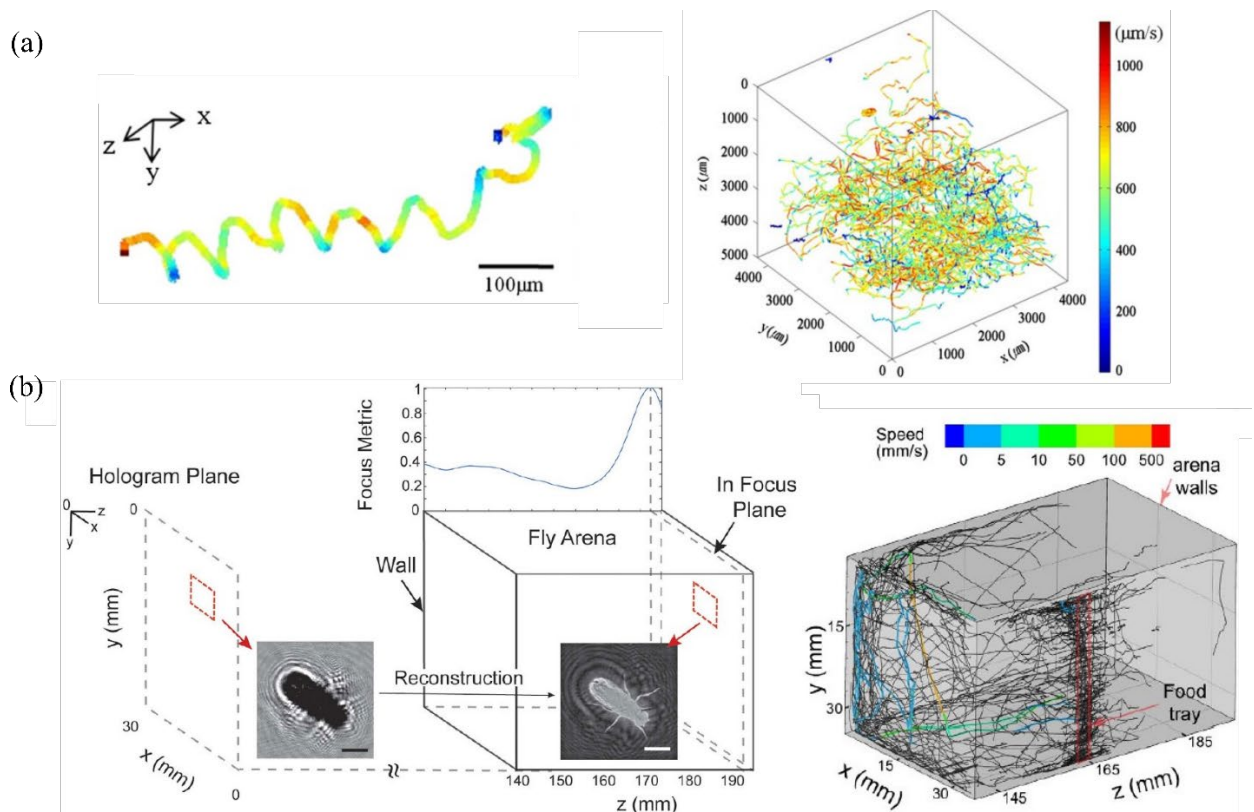

Fig. 6. (a) Examples of 3D trajectories of a single and group of phytoplankton obtained using DIH with sharpness-based focus metrics. Adapted from [83] Copyright (2024), with permission from Elsevier. (b) Detection of in-focused plane of a drosophila and 3D trajectories of a group of drosophilae using DIH with L1-norm based focus metrics. Adapted from [93].

Despite their widespread use, traditional multi-step approaches have several limitations. The numerical reconstruction and focus evaluation across multiple depths are time-consuming, leading to significant computational load. These methods may suffer from limited depth accuracy, impacting the precision of particle localization along the longitudinal axis. The field of view can be restricted due to the border effect. The border effect leads to distorted and low-contrast images in the reconstruction planes and restricts the usability of traditional multi-step methods that rely on the reconstruction approach. Additionally, the presence of spurious twin images resulting from the holographic reconstruction process [99] can interfere with accurate localization and segmentation. To overcome these challenges, alternative methods such as inverse problem-solving approaches and ML techniques have been developed. These methods offer improvements in computational efficiency, accuracy, and the ability to handle complex flow scenarios, addressing many of the limitations inherent in the traditional multi-step approach.

## 4.2. Inverse method

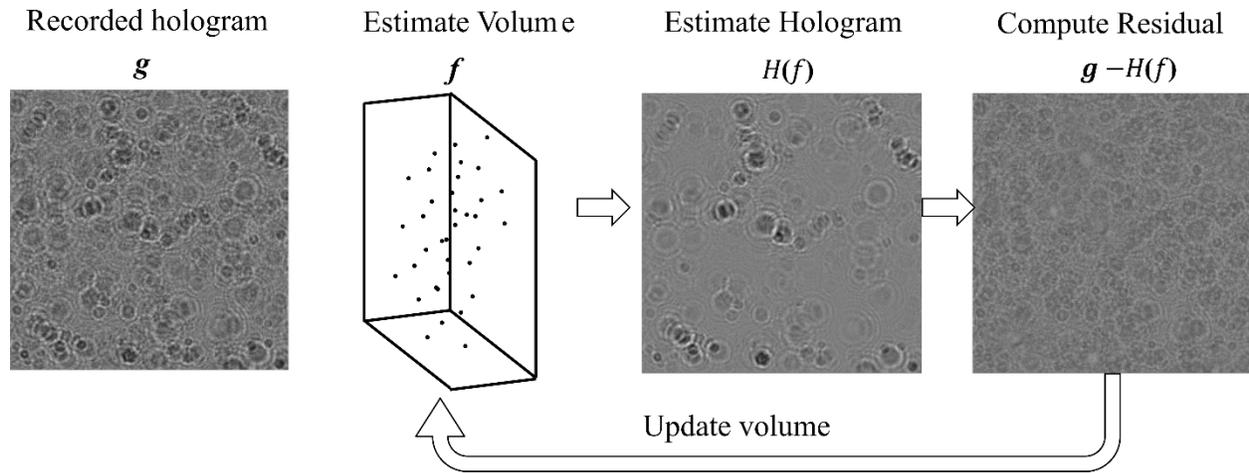

**Fig. 7.** Schematics illustrating the iterative process of implementing inverse method for hologram processing.

The inverse problem approach offers an alternative to traditional hologram reconstruction methods by directly retrieving the 3D characteristics of the particle field through iterative refinement of a parametric model. Instead of simulating diffraction to reconstruct the hologram, this method aims to minimize the discrepancy between the actual recorded hologram and a simulated hologram generated from an estimated particle distribution. The process involves creating an initial model of the particle field, comprising guessed positions, sizes, and other properties, and then simulating the hologram that such a distribution would produce. By comparing this simulated hologram with the real one, the model parameters are iteratively adjusted until the simulated hologram closely matches the recorded hologram. Mathematically, the recorded hologram $g$ can be modeled using a parametric function $H(f)$ based on a diffraction model:

$$H(f) = I_0 - \sum_{k=1}^{n} \alpha_k h_k \quad (8)$$

where $n$ is the total number of particles, $I_0$ represents the incident intensity, $\alpha_k$ is the amplitude factor of the $k$-th particle, and $h_k$ is the diffraction pattern of the $k$-th particle. The inverse problem is formulated as an optimization task, seeking the optimal set of particle parameters $\{x_k, y_k, z_k, r_k\}_{k=1...n}$ (positions and radii) that minimize the difference between the recorded hologram $g$ and the estimated hologram $H(f)$. This optimization can be expressed as:

$$\hat{f} = \arg \min_{f} \{\|g - H(f)\|_2^2 + \lambda R(f)\} \quad (9)$$

where $\|\cdot\|_2^2$ denotes a squared Euclidean term, $\lambda R(f)$ is the regularization term, and $\lambda$ is the regularization parameter balancing data fidelity and regularization. The inverse problem is commonly solved iteratively as illustrated in Fig. 7.

Soulez et al. (2007) pioneered one of the earliest inverse formulations for holographic reconstruction, utilizing four-dimensional parametric optimization to estimate the 3D locations and radii of spherical particles with high precision and robustness to noise [100]. Their iterative particle detection algorithm refines the particle parameters $\{x_k, y_k, z_k, r_k\}$ at each step by minimizing the cost function defined in Eq. (9). While inverse problems commonly use explicit regularization techniques (e.g., L1-norm or L2-norm) to enforce sparsity or smoothness, the approach described in Soulez et al. (2007) instead focuses on reducing the difference between the hologram model and the observed data [100]. They incorporate statistical weighting of pixel contributions based on noise variability, which partially addresses the ill-posed nature of the problem but does not constitute formal mathematical regularization. Rather than explicitly penalizing model complexity or imposing parametric constraints, the algorithm relies on iterative adjustments of residual images and localized optimization to stabilize and refine the solution. Because there is no explicit regularization term, the computational time scales linearly with the number of particles, resulting in significant processing demands. This limitation renders the method unsuitable for fluid flow applications involving thousands of particles across multiple frames. Moreover, the approach becomes inefficient in high particle concentration scenarios and is restricted to spherical particles, further limiting its broader applicability.

Brady et al. (2009) advanced the inverse problem approach by formulating it as a sparsity-constrained global optimization problem [101]. They introduced total variation (TV) regularization $R(f) = \|\nabla f\|_1$, promoting smoothness in the reconstructed volume while preserving sharp edges. This method effectively captures fine details and complex 3D particles with irregular shapes or varying sizes. Despite its strengths, the method is computationally intensive and depends on the assumption of smoothness, which may not hold for all particle distributions. To enhance computational efficiency and generality, Mallery and Hong (2019) combined sparsity and smoothness by employing fused lasso regularization $R(f) = \|\nabla f\|_1 + \|f\|_1$ [102]. This approach allowed them to handle larger volumes, more complex particle shapes (Fig. 8a), and high-noise environments. However, scalability remained a challenge due to substantial memory requirements and the necessity for meticulous tuning of regularization parameters.

A recent development in holographic imaging is the introduction of 3D differentiable holography ($\partial H^3$) by Wu et al. (2024) [103]. This method addresses the inherent limitations of traditional approaches in high-concentration volumetric particle imaging by integrating a more accurate forward propagation model with automatic differentiation techniques. Unlike earlier methods that rely on linear approximations suitable for weakly scattering objects, $\partial H^3$ employs a nonlinear multislice beam propagation model to account for multiple scattering phenomena in dense particle fields. This advanced model more closely mirrors real-world physical processes, thereby enhancing the fidelity of volumetric reconstructions.

While traditional inverse methods focus on directly minimizing the discrepancy between the recorded hologram and a simulated one through iterative refinement of a parametric model (as described in Eq. 9), alternative approaches have been developed to address specific challenges in holographic particle characterization. Lee et al. (2007) introduced a method that diverges from the conventional inverse problem framework by utilizing Lorenz-Mie scattering theory [104]. Instead of performing full optimization, they fit a theoretical scattering model to the holographic data to extract particle characteristics such as size, refractive index, and 3D positions of colloidal particles. This approach provides valuable insights into particle properties by directly interpreting the scattering patterns without iterative minimization. In contrast, Chen et al. (2021) proposed an advanced inverse method that extends beyond particle reconstruction to include flow velocity estimation [105]. Their joint optimization framework simultaneously reconstructs 3D particle volumes and flow velocities by formulating the problem as a coupled inverse problem. This is achieved by incorporating flow velocities as a regularization prior in the particle volume reconstruction and iteratively refining flow estimations based on the updated particle positions. By including domain-specific priors such as particle sparsity and flow smoothness, along with Tikhonov regularization, they significantly enhance reconstruction quality. This method proves highly effective even in high-noise environments and is suitable for large-scale fluid flows with high concentration particle fields (Fig. 8b).

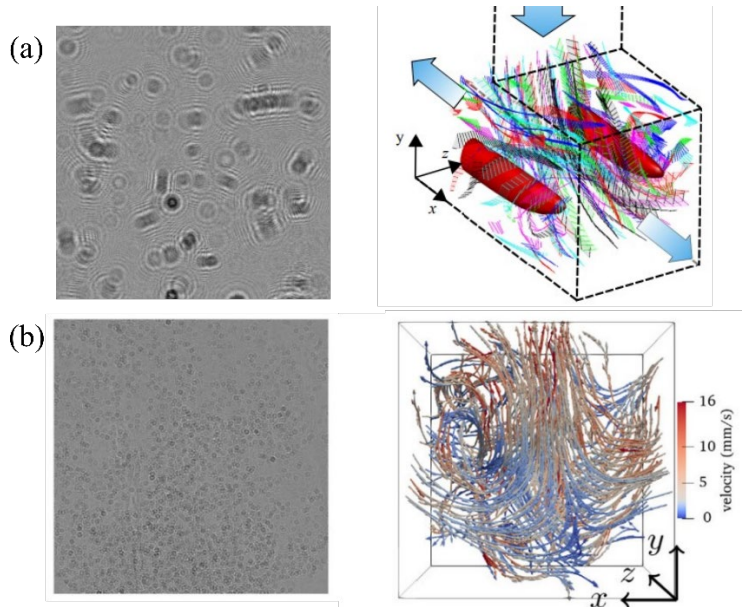

**Fig. 8.** (a) Sample hologram and the corresponding reconstructed 3D trajectories and orientations of microfibers moving in a T-junction flow using inverse method with fused lasso regularization. Adapted with permission from [102] © Optical Society of America. (b) Sample hologram and the corresponding reconstructed 3D trajectories of tracers in a flow field using the inverse method based on a joint optimization approach. Adapted from [105].

While inverse problem approaches offer significant advantages in terms of accuracy and the ability to handle complex particle distributions, they come with challenges. The computational intensity of iterative optimization, especially in high-concentration particle fields, can be substantial. Memory requirements increase with the volume size and particle count, potentially limiting scalability. Additionally, the success of these methods often hinges on the choice of regularization parameters and the accuracy of the forward model. Misalignments between the model and actual physical processes can lead to suboptimal reconstructions. Nevertheless, the inverse problem approach represents a powerful tool in holographic particle tracking, enabling precise reconstruction of particle fields and flow characteristics that are difficult to achieve with traditional methods. Continued advancements in computational algorithms and hardware are likely to mitigate current limitations, further enhancing the applicability of these methods in complex flow diagnostics and particle tracking applications.

## 4.3. Machine learning based approaches

The integration of ML into DH has revolutionized hologram processing by addressing many of the limitations associated with traditional methods. Conventional techniques often involve manual parameter tuning and are computationally intensive, especially when handling complex tasks such as image reconstruction, classification, feature segmentation, and phase recovery. Machine learning algorithms streamline these processes by automatically learning patterns and features directly from holographic data, reducing the need for manual intervention and enabling faster, more efficient processing. This advancement has extended the capabilities of DH beyond fundamental research, enabling real-time applications across various sectors, including manufacturing, environmental monitoring, medical diagnostics, and more.

In the measurements of flow field and particle dynamics, ML approaches can be broadly classified into two categories: multi-stage models and single-stage models. Multi-stage models utilize separate ML algorithms for different stages of hologram processing, such as particle detection and localization. These models often replace specific steps in traditional multi-step methods with ML algorithms to improve accuracy and efficiency. One prominent example is the application of the "You Only Look Once" (YOLO) object detection framework. Known for its real-time processing and high accuracy, YOLO has been adapted to detect various particles directly from holograms. Researchers have successfully employed YOLO-based models to identify colloidal particles [106], dental aerosols [107], spray droplets [108], swimming microorganisms [109], snow particles [110], and different species of yeast and plankton [111]. Additionally, Support Vector Machines (SVMs) have been used to distinguish microplastics from other particles like marine diatoms [112]. While these detection models offer significant improvements over traditional methods, they can be sensitive to noise and may struggle with high particle concentrations. Overlapping interference patterns in densely populated holograms can complicate detection and reduce accuracy.

Accurate particle localization is essential for applications that require tracking particles over time. Machine learning models have been developed to predict the focus depth of particles directly from holograms, eliminating the need for numerical reconstruction. This is often formulated as a classification or regression problem. Classification-based models utilize convolutional neural networks (CNNs) to categorize hologram images into discrete focus levels, effectively determining the in-focus depth of particles. Ren et al. (2018) [113], Jaferzadeh et al. (2019) [114], and Pitkäaho et al. (2019) [115] developed CNN architectures that predict the focus plane directly from hologram amplitude images without performing numerical propagation or fitting to diffraction theories. These networks are trained to recognize patterns associated with sharp features in the hologram, indicating particles in focus. By learning these patterns, the models can rapidly classify new holograms and identify the depth at which particles are sharply imaged. In particular, Lee et al. (2019) integrated the circular Hough Transform (CHT) for microparticles detection with a CNN to accurately classify the segmented holograms into different depth classes, achieving a depth estimation accuracy that was nearly three times more precise than conventional methods (Fig. 9a) [116].

For practical scenarios demanding continuous depth estimation, regression models are more suitable. Wu et al. (2018) [117] and Altman and Grier (2023) [106] employed deep neural networks comprising convolutional layers to extract features from holographic images and fully connected layers to map these features to quantitative particle properties such as 3D position, size, and refractive index. Huang et al. (2021) demonstrated the effectiveness of integrating a convolutional recurrent neural network (RNN) within a Generative Adversarial Network (GAN) framework to perform both autofocusing and reconstruction of human tissue holography images [118]. Their approach processes multiple holograms captured at different sample-to-sensor distances simultaneously, enhancing depth estimation accuracy. Additionally, Ren et al. (2018) showcased the precision of a CNN regression model in depth prediction for amplitude and phase-only objects imaged at multiple distances [119].

These regression models offer a high degree of accuracy and, once trained, can predict the in-focus depth in real time. This capability makes them highly efficient for applications such as microscopy, autofocus systems, and microfluidic studies (Fig. 9). However, their performance tends to diminish in scenarios with higher particle concentrations due to challenges like overlapping or truncated interference patterns in the holograms. The overlapping diffraction patterns make it difficult for the models to extract distinct features associated with individual particles, leading to reduced localization accuracy. Additionally, these models face challenges with boundary images, where particles located near the edges of the holographic field may not be accurately detected due to partial visibility or edge effects.

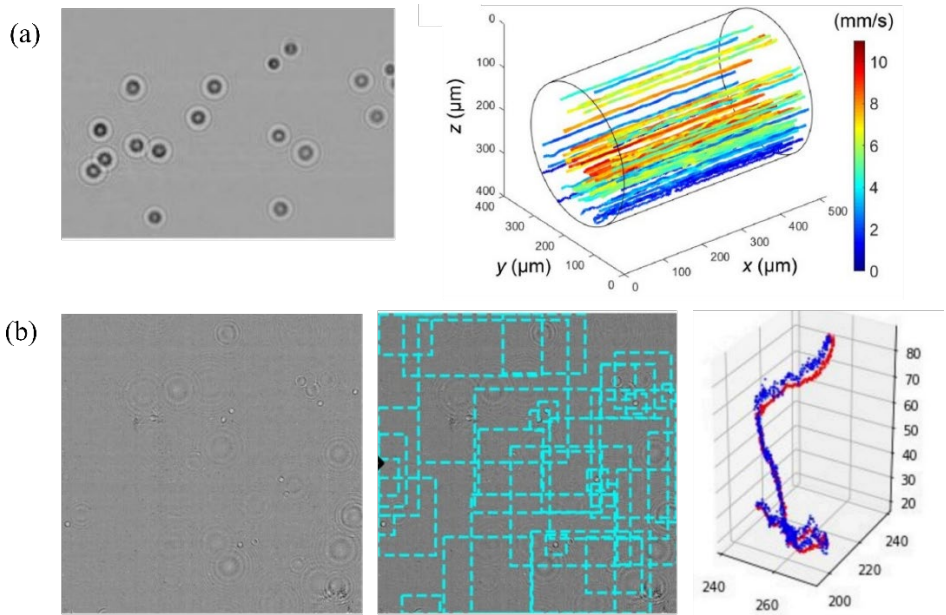

**Fig. 9.** (a) Sample hologram and the corresponding reconstructed 3D trajectories of microparticles in a microtube flow using the circular Hough Transform for particle detection, followed by a CNN for particle localization. Adapted from [116] Copyright (2024), with permission from Springer Nature. (b) Sample hologram, the corresponding bounding box prediction and sample tracks of swimming E. Coli generated using DH with YOLO-based ML method. Adapted from [109].

Single-stage models represent a significant advancement by performing particle detection, localization, and segmentation within a single neural network. These models aim to streamline the processing pipeline, reducing computational load and improving efficiency. An example of this approach is OSNet, introduced by Zhang et al. (2022) [120]. Inspired by the YOLO framework, OSNet processes 2D holograms and directly outputs the 3D coordinates of particles, effectively bypassing the need for numerical reconstruction. This method accelerates processing but may experience performance degradation in conditions with high particle concentration or significant noise unless supplemented with additional computational resources.

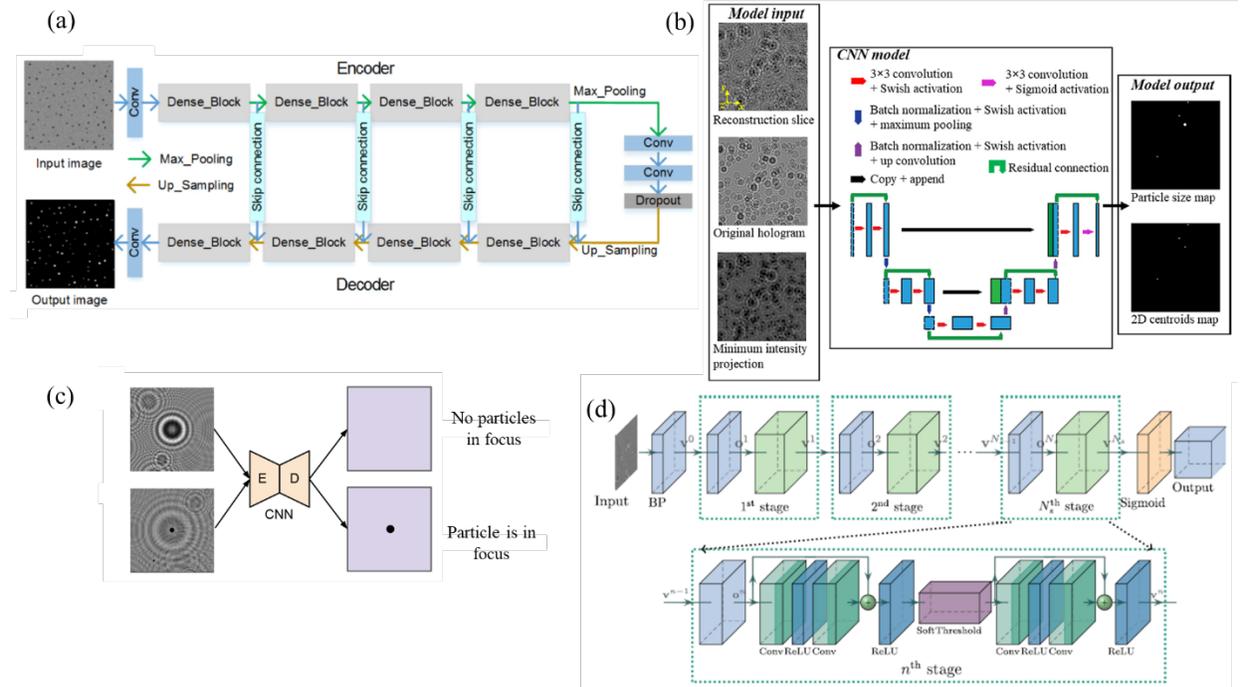

**Fig. 10.** Different single-stage models used for hologram processing: (a) U-Net model adapted from [122] Copyright (2024), with permission from Elsevier, (b) U-Net model adapted from [123] Copyright (2024), with permission from Elsevier, (c) encoder-decoder network (HolodecML) adapted from [125], and (d) MB-HoloNet adapted from [126].

To address the challenges posed by higher particle concentrations, architectures like U-Net have been employed [95,121-124]. U-Net's encoder-decoder structure with skip connections allows it to efficiently handle both sparse and dense data by preserving important spatial information throughout the network (Fig. 10a). This makes U-Net particularly effective in complex datasets where traditional methods falter due to increased cross-interference from densely packed particles. Enhancements like residual connections within U-Net architectures further improve training stability and accuracy in high particle concentration scenarios (Fig. 10b) [123]. Applications of U-Net-based models include agricultural spray analysis [49,123] and the study of interactions between water droplets and swirling air currents [124]. Schreck et al. (2022) developed a similar encoder-decoder network that reframes the problem as bounding box regression, directly predicting particle coordinates (Fig. 10c) [125]. While this approach is computationally efficient, it also faces challenges when dealing with very high particle concentrations. To enhance model adaptability across different holographic conditions, such as varying particle sizes, shapes, concentrations, and optical setups researchers have explored incorporating physical principles into ML models. Chen et al. (2021) introduced MB-HoloNet, a physics-informed neural network that reconstructs 3D particle fields from 2D holograms by learning parameters defining the point spread function (PSF) (Fig. 10d) [126]. By integrating knowledge of the optical system and diffraction patterns into the learning process, MB-HoloNet improves generalizability and reduces reliance on extensive labeled datasets.

Despite significant advancements, ML models for hologram processing still face several challenges. A common issue is their sensitivity to noise and diminished performance in environments with high particle concentrations. In densely populated holograms, overlapping interference patterns complicate feature extraction, making it difficult for models to accurately detect and localize individual particles. Single-stage models often require substantial computational resources to maintain performance under noisy or densely populated conditions, which may not be practical for all applications. Additionally, many ML models rely heavily on extensive labeled datasets for training, which can be time-consuming and labor-intensive to produce. This dependency limits the models' generalizability to varying holographic conditions and different experimental setups. While incorporating physical principles into ML models has shown promise in enhancing their generalizability, these physics-informed models are still in the early stages of development and face challenges related to scalability and practical implementation.

We have summarized the review of hologram data processing in Table.1. This table provides a structured overview of various data processing methods used in hologram analysis, including traditional multi-stage approaches, inverse methods, and emerging machine learning-based solutions. Each method's key strengths—such as accuracy, speed, generalizability, computational efficiency, and robustness to noise—are presented alongside its associated limitations, which may include computational intensity, reduced effectiveness at high particle concentrations, susceptibility to noise and overlapping particles, and scalability challenges.

Table 1. Summary of different digital hologram processing methods for 3D particle tracking and flow diagnostics.

| | Sub-class | | Advantages | Limitations | References |
|---|---|---|---|---|---|
| **Traditional multi-stage approaches** | | | | | |
| Reconstruction | Conventional reconstruction | | - Computed directly via convolution of diffraction kernel | - Low signal-to-noise ratio, especially in high particle concentration<br>- Artifacts introduced by assumptions in the reconstruction kernel | [35-39] |
| Reconstruction | Deconvolution | | - Improves the signal-to-noise ratio<br>- Facilitates subsequent particle localization | - High computational cost<br>- Artifacts tied to the choice and accuracy of the deconvolution kernel | [40-43,76,77] |
| Particle localization | Longitudinal localization using focus metrics | Spatial based | - Localized analysis reduces interference from distant particles<br>- Well-suited for particles with distinct edges<br>- Computationally efficient | - Susceptible to noise in high particle concentrations<br>- Less effective for transparent particles<br>- Prone to optical distortions | [78-91] |
| Particle localization | Longitudinal localization using focus metrics | Spectral based | - Avoids repeated image reconstruction at multiple depths<br>- Potentially efficient when spectral data is well-defined | - Struggles with noise or cross-interference that skews the spectral content<br>- Limited effectiveness for complex morphologies or high concentrations | [92-94] |
| Particle localization | Longitudinal localization using focus metrics | Wavelet based | - Robust against noise by separating frequency components | - Intensive computation required for multi-scale wavelet decomposition | [96] |
| Particle localization | 3D localization using 3D segmentation | | - Effective for high particle concentration fields<br>- Direct 3D measurement without separate focusing | - Accuracy drops for irregular particle clusters or strong cross-interference<br>- High computational cost | [97,98] |
| **Inverse methods** | | | | | |
| Regularization | Weighting of pixel contribution | | - Provides high precision and robustness against noise | - Consumes significant computation time | [100] |
| Regularization | Total variation (TV) regularization | | - Captures fine details and complex 3D particles with irregular shapes or varying sizes | - Demands intensive computational effort<br>- Assumption of smoothness may not hold for all the particle concentrations | [101] |
| Regularization | Fused lasso regularization | | - Handles larger volumes and more complex particle shapes | - Encounters scalability challenges<br>- Requires meticulous tuning of regularization parameters | [102] |
| **Machine learning based approaches** | | | | | |
| Multi-stage network | Particle detection | YOLO | - Enables real-time processing<br>- Delivers high detection accuracy | - Struggles under high particle concentrations and overlapping diffraction patterns | [106-111] |
| Multi-stage network | Particle detection | SVMs | - Ensures high classification accuracy | - Exhibits sensitivity to noise and overlapping patterns | [112] |
| Multi-stage network | Particle localization | Classification models | - Provides precise depth estimation<br>- Offers rapid classification speed | - Shows decreased performance under high particle concentrations | [113-116] |
| Multi-stage network | Particle localization | Regression models | - Delivers improved prediction accuracy<br>- Supports real-time capability | - Suffers from reduced localization accuracy in dense particle scenarios<br>- Faces challenges with boundary images | [106,117-119] |
| Single stage network | | OSNet | - Offers high processing speed<br>- Provides accurate 3D coordinate estimation | - Experiences performance degradation under high particle concentrations or significant noise | [120] |
| Single stage network | | U-Net | - Delivers improved accuracy for high concentration particle holograms | - Depends entirely on data-driven methods, reducing generalizability | [49,95,121-124] |
| Single stage network | | MB-HoloNet | - Promotes improved generalizability across diverse particle conditions<br>- Reduces reliance on extensively labeled datasets | - Remains in early development stage<br>- Encounters scalability issues | [126] |

## 5. Summary

Digital holography (DH) has emerged as a powerful and versatile tool for three-dimensional (3D) flow field and particle tracking, providing unprecedented insights into complex fluid phenomena and particle dynamics. In this review, we thoroughly investigated the core principles of DH and examined various hardware configurations and data processing techniques that are crucial for extracting meaningful 3D particle information from recorded holograms.

Among the hardware configurations discussed, digital inline holography (DIH) stands out for its simplicity and cost-effectiveness, leveraging a straightforward optical arrangement that aligns the object and reference waves along the same axis. This configuration is particularly advantageous for applications requiring minimal equipment and ease of alignment. We have also reviewed modifications that improve longitudinal resolution and mitigate artifacts, such as employing Mach–Zehnder setups, introducing dual inline holograms with slight focal plane separation, or adding astigmatism via cylindrical lenses. Meanwhile, off-axis and dual- or multiple-view holography setups enhance depth resolution and suppress twin images, albeit at the cost of increased complexity, stricter alignment requirements, and higher overall expenses.

In parallel, hologram data processing techniques have evolved to meet the demands of increasingly complex particle and flow fields. We have reviewed a variety of data processing approaches, including traditional multi-step methods, inverse methods, and, more recently, machine learning (ML)-based techniques. Traditional multi-step methods rely on numerically reconstructing the optical field and then applying focus metrics to localize particles in 3D space. In contrast, inverse methods bypass the explicit field reconstruction step and instead treat hologram processing as an optimization problem, directly extracting 3D particle characteristics by minimizing discrepancies between recorded and simulated holograms. ML-based techniques have further advanced the field by automating particle detection, localization, and segmentation, often achieving higher throughput and enhanced robustness compared to conventional approaches. While most ML models employ multiple specialized stages, for instance, separate modules for particle detection and focus prediction, emerging single-stage frameworks aim to integrate these tasks within a single network architecture. In particular, U-Net and other encoder–decoder models have been adapted for hologram processing, demonstrating their potential to handle high particle concentrations effectively.

In summary, we have examined a range of digital holography configurations and data processing algorithms developed over the years. Notable progress in both non-ML and ML-based techniques has substantially improved particle tracking and flow diagnostics in research settings, and these advancements hold promise for broader application in industrial environments and other complex scenarios. As these methodologies continue to evolve, DH is poised to become even more accessible, reliable, and capable of capturing the intricate details of fluid flows and particle dynamics with unprecedented accuracy and efficiency.

## 6. Current limitations and challenges

Despite significant advancements made in DH as outlined previously, several inherent challenges continue to constrain its widespread industrial and practical adoption. These challenges primarily revolve around a hardware–software trade-off: while DIH's simplicity is appealing, it can limit axial resolution and increase susceptibility to noise and artifacts. More complex hardware setups, such as off-axis or multi-view holography, can mitigate some of these issues but come at the expense of greater complexity, cost, and alignment sensitivity. Consequently, a substantial portion of recent research has focused on developing advanced data processing methods that leverage the simplicity of DIH while counteracting its drawbacks. Despite notable progress, these approaches still face a variety of technical hurdles, as detailed below.

## 6.1 Hardware-Related Trade-Offs

DIH's core strength lies in its simple, collinear configuration, which reduces both costs and alignment complexity. However, this arrangement inherently places the object and reference beams along the same axis, making it more challenging to separate them effectively and accurately determine particle positions along the depth (axial) direction. High particle concentrations exacerbate these issues, as increased speckle noise, cross-interference, and twin-image artifacts can further degrade measurement accuracy. In contrast, off-axis and multi-view configurations improve depth resolution and suppress twin images but demand additional optical components and more stringent alignments, raising both complexity and expense. This balancing act creates a dilemma in hardware selection: one can choose a DIH-based setup that is compact and economical but prone to artifacts, or invest in a more complex system that offers higher fidelity at significantly greater cost and complexity.

## 6.2 Challenges in Non-ML Data Processing

To compensate for the inherent limitations of DIH hardware, an array of non-ML data processing techniques has been developed [19, 34, 99]. These methods—ranging from refined focus metrics and deconvolution strategies to inverse approaches—aim to enhance axial localization, improve signal quality in noisy conditions, and alleviate twin-image artifacts. Although effective in many scenarios, these solutions often require extensive parameter tuning that can be highly scenario dependent. For example, adjustments to focus thresholds, regularization parameters, or deconvolution kernels can be labor-intensive, and identifying optimal settings is not always straightforward. Additionally, conventional (non-ML) techniques are frequently computationally expensive, prolonging processing times and making real-time feedback challenging. This limitation is especially critical when analyzing large, complex datasets, such as those associated with turbulent flows, sprays, or combustion environments [33], where rapid, on-the-fly data interpretation would be most beneficial.

## 6.3 Constraints of Data-Driven Methods

Machine learning and deep learning (DL) approaches have recently attracted significant attention as tools to reduce manual intervention and accelerate computational speed [127]. By learning from labeled examples, these algorithms can automate key steps in data processing, mitigating some of the burdens of parameter tuning. However, current ML/DL models depend heavily on their training datasets, which must represent a diverse range of particle sizes, morphologies, optical conditions, and noise levels. Even small deviations from training conditions can degrade performance, undermining the models' robustness and generalizability. Moreover, while ML/DL methods can speed up computations, they have yet to deliver truly instantaneous, real-time 3D reconstructions—an essential capability for dynamic or industrial applications that demand immediate feedback or control.

In summary, DH, and DIH in particular, currently faces a multifaceted set of challenges spanning hardware design, data processing, and ML-driven approaches. Although advances in algorithms have partially alleviated some of DIH's inherent shortcomings, the reliance on manual tuning and computationally intensive methods remains problematic. The emerging application of ML/DL offers promise but introduces new dependencies on training data and still falls short of delivering immediate, real-time solutions. Addressing these limitations will be crucial for unlocking DH's full potential for reliable, automated, and high-throughput 3D particle diagnostics, ultimately guiding the field toward more robust, physics-informed ML frameworks and faster, more adaptive data processing algorithms.

## 7. Future Prospect

As DH advances toward more robust, accurate, and real-time 3D particle diagnostics, addressing the limitations highlighted in the previous sections will be critical. By integrating physics-based models into ML frameworks, exploring advanced architectures, and embracing adaptive and online learning strategies, DH can significantly expand its applicability and performance across various fields. The following subsections outline several key directions and opportunities for future development.

### 7.1 Enhancing Generalizability Through Physics-Informed and Foundation ML Models

A promising avenue for enhancing the generalizability and robustness of ML in DH is the incorporation of physics-based constraints. Physics-Informed Neural Networks (PINNs) embed physical laws into the learning process, potentially reducing reliance on large, scenario-specific training datasets. By capturing fundamental fluid and optical principles, these models can adapt to variations in particle size, morphology, and concentration, as well as diverse optical conditions. Although refining the underlying physics to fully reflect real-world complexities can increase computational costs [128], improved physical approximations and hybrid approaches can bolster the model's realism and scope of application.

In parallel, the concept of "foundation models" trained on diverse and extensive datasets [129] shows promise for DH. Such models could handle multiple tasks—particle detection, localization, reconstruction, and noise reduction—under one comprehensive framework. By streamlining workflows and reducing computational overhead, foundation models can facilitate rapid adaptation to new experimental setups and conditions, expediting both research and industrial deployments.

### 7.2 Improving Computational Efficiency with Advanced Architectures

Improving computational efficiency is essential for achieving fast feedback and control. Generative Adversarial Networks (GANs) and Variational Autoencoders (VAEs) have already demonstrated their potential in accelerating data processing and improving image quality. GANs can synthesize large, realistic training holograms from limited data [130,131], helping alleviate data bottlenecks. They also enable faster reconstruction with extended depth of field, facilitating dynamic observations under complex flow conditions [132,133]. Autoencoders and their variants capture complex nonlinear relationships without extensive labeled data, improving phase reconstruction accuracy and reducing manual interventions [134,135]. The emergence of Vision Transformers (ViTs) offers another powerful option for image reconstruction tasks, improving both speed and fidelity [136]. By combining these advanced architectures, future DH systems can drastically reduce computational costs, making continuous, on-the-fly diagnostics more accessible.

### 7.3 Elevating Accuracy in Dense Particle Environments

Achieving high accuracy in dense particle environments remains challenging, as twin-image artifacts and heavy speckle noise can obscure particle features. Recent advancements in ML architectures have shown promise in this regard. GANs can help suppress twin-image artifacts and improve overall image clarity, enhancing precision in particle tracking and flow diagnostics [137]. When coupled with physical constraints, PINNs can further refine phase imaging and segmentation quality, even under challenging conditions involving dense particle concentrations [126, 138]. These developments will bolster the applicability of DH, enabling it to handle a wider spectrum of flows and environments with improved reliability.

### 7.4 Embracing Adaptive and Online Learning Strategies

As measurement conditions evolve—due to changing flow regimes, shifting particle concentrations, or variations in optical alignment—adaptive learning methods will become increasingly important. Online and incremental learning approaches can allow ML models to update and improve in real time [139], ensuring that DH systems remain accurate, efficient, and robust despite changing scenarios. Moreover, exploring novel architectures such as graph neural networks or spiking neural networks may offer fresh avenues for modeling complex holographic data. Hybrid models that integrate GANs with ViTs or PINNs can push performance boundaries even further, paving the way for scalable, flexible, and future-proof DH solutions.

### 7.5 Broadening Applications Across Various Fields

With improvements in generalizability, computation, and accuracy, DH has the potential to revolutionize a wide range of applications. In manufacturing, particularly within the pharmaceutical industry, DH can facilitate real-time, in-situ process monitoring, supporting Process Analytical Technology (PAT) initiatives and quality control [140]. Beyond pharmaceuticals, DH can be employed in the food and beverage sector to monitor microbial processes [141], enable real-time bacteria detection in sterile liquid products [142], and assist in the semiconductor industry by detecting and characterizing particulate contaminants [143].

In environmental monitoring, DH can deliver crucial insights into pollen dispersion [30], smoke aerosol [144] and microplastic transport [112], pesticide spray drift [49], harmful algal bloom [145], larval fish intake into power plants [146], snow particle settling [110], cloud microphysics [73], and indoor air quality [147]. Specifically, in air quality monitoring, to achieve precise and reliable measurements, deploying high-fidelity sensors is essential [148]. These sensors, when integrated with DH capabilities, provide real-time particulate data, bridging the gap between traditional methods and advanced monitoring systems. DH can significantly enhance these systems, aiding pollution control efforts and contributing to broader environmental stewardship goals.

In biology and medicine, the ability of DH to track cells and microorganisms at high spatial and temporal resolutions opens the door to advanced diagnostic and therapeutic applications. Early detection of circulating tumor cells (CTCs) [149], rare cell isolation [150], sperm motility analysis [27], yeast cell metabolic state monitoring [111], and biofilm formation studies [151] represent just a few of the many avenues where DH can provide valuable, minimally invasive insights.

In conclusion, DH has the potential to revolutionize the measurement and analysis of 3D particle dynamics and flow fields. By overcoming current limitations through the development of more generalizable and accurate ML models, computationally efficient architectures, and adaptive learning strategies, DH is positioned to deliver robust, precise, and real-time diagnostics across a wide range of applications. These advancements will not only enhance our understanding of complex fluid systems but also expand DH's influence beyond research laboratories, driving breakthroughs in industrial processes, environmental monitoring, and biomedical sciences. As the field progresses, DH is set to become an indispensable technology for advancing scientific discovery and addressing real-world challenges in particle and flow diagnostics.

## Data availability statement

No new data were created or analyzed in this study.

## Acknowledgments

The authors acknowledge the assistance provided by Winfield United, Wisconsin, USA, to Shyam Kumar M during his postdoctoral research.The authors acknowledge the assistance provided by Winfield United, Wisconsin, USA, to Shyam Kumar M during his postdoctoral research.